\documentclass[sigconf, nonacm]{acmart}

\newcommand\vldbdoi{XX.XX/XXX.XX}
\newcommand\vldbpages{XXX-XXX}
\newcommand\vldbvolume{14}
\newcommand\vldbissue{1}
\newcommand\vldbyear{2024}
\newcommand\vldbauthors{\authors}
\newcommand\vldbtitle{\shorttitle} 
\newcommand\vldbavailabilityurl{}
\newcommand\vldbpagestyle{plain}

\newcommand{\sys}{\mbox{\textsc{SketchQL}}\xspace}

\newcommand{\paperTitle}{\sys Demonstration: Zero-shot Video Moment Querying with Sketches}

\newcommand{\miris}{\mbox{\textsc{Miris}}\xspace}

\newcommand{\sketcher}{\mbox{\textsc{Sketcher}}\xspace}
\newcommand{\matcher}{\mbox{\textsc{Matcher}}\xspace}
\newcommand{\tuner}{\mbox{\textsc{Tuner}}\xspace}

\usepackage{anyfontsize}
\usepackage[hyphens]{url}
\usepackage{pifont}

\definecolor{linkcolor}{HTML}{647382}
\definecolor{citecolor}{HTML}{647382} %
\definecolor{urlcolor}{rgb}{0.4,0.2,0.2}
\definecolor{sqlcolor}{HTML}{965d67}
\definecolor{smtcolor}{HTML}{5d968c}
\definecolor{webblue}{rgb}{0,0,.7}
\definecolor{webgreen}{rgb}{0,.5,0}
\definecolor{webbrown}{rgb}{.6,0,0}
\definecolor{notecolor}{HTML}{FFF8DC}

\usepackage{amsmath,amsopn,amssymb}
\usepackage[cal=boondoxo]{mathalfa}
\usepackage{subcaption}
\usepackage{endnotes,microtype,xspace,graphicx,fancyvrb,multirow}
\usepackage{supertabular,booktabs}
\usepackage{array,underscore, relsize}
\usepackage{fancyhdr}
\usepackage{enumitem}
\usepackage{balance}
\usepackage{booktabs}
\usepackage{pifont}
\usepackage{listings}
\usepackage{paralist}

\usepackage{multirow}
\usepackage[scaled]{beramono}
\usepackage{tabularx}
\usepackage{multirow}

\makeatletter
\newcommand\BeraMonottfamily{%
  \def\fvm@Scale{0.85}
  \fontfamily{fvm}\selectfont
}
\makeatother

\definecolor{mymauve}{rgb}{0.58,0,0.82}

\lstdefinestyle{SQLStyle}{
  language=SQL,
  basicstyle={\footnotesize\ttfamily},
  breaklines=true,
  frame=none,
  numbers=none,
  keepspaces=true,
  captionpos=b,
  stringstyle=\color{mymauve},
  keywordstyle=\color{blue},
  commentstyle=\color{dkgreen},
}

\lstdefinestyle{PyStyle}{
  language=Python,
  basicstyle=\ttfamily\scriptsize, 
  aboveskip = 0.05in,
  belowskip = 0.05in,
  breaklines=true,
  float=tp,
  floatplacement=tbp,
  frame=none,
  numbers=none,
  keepspaces=true,
  captionpos=b,
  showstringspaces=false,
  emph={MyClass,__init__},          
  stringstyle=\color{deepgreen},
  emphstyle=\ttb\color{deepred},    
  keywordstyle=\color{blue},
  commentstyle=\color{webgreen},
  morekeywords={self,def, for, sum, in, and}
}

\lstdefinestyle{ScriStyle}{
language=SQL,
basicstyle=\BeraMonottfamily\footnotesize, 
keywordstyle=\color{smtcolor}\bfseries,
morekeywords={and, or, not},
aboveskip = 0.05in,
belowskip = 0.05in,
literate = {-}{-}1, 
}

\usepackage{aliascnt}

\usepackage[ruled,linesnumbered,vlined]{algorithm2e}
\usepackage{setspace}
\usepackage{enumitem}

\SetAlFnt{\small}
\SetAlCapFnt{\small}
\SetAlCapNameFnt{\small}

\usepackage{semantic}

\usepackage{stmaryrd}
\usepackage{ltablex}
\usepackage{mathtools}
\usepackage{adjustbox}
\usepackage[group-separator={,}]{siunitx}
\usepackage[capitalize,noabbrev,nameinlink]{cleveref}

\crefname{lstlisting}{listing}{listings}
\Crefname{lstlisting}{Listing}{Listings}

\usepackage{chngcntr}
\counterwithout{equation}{section}

\newcommand{\hide}[1]{}

\newcommand{\PP}[1]{
\vspace{2px}
\noindent{\bf\textsc{#1}.}\xspace
}

\newcommand{\PPP}[1]{
\vspace{0.05in}
\noindent{\textit{\IfEndWith{#1}{.}{#1}{#1.}}}
}

\newcommand{\squishitemize}{
 \begin{list}{$\bullet$}
  { \setlength{\itemsep}{0pt}
     \setlength{\parsep}{0pt}
     \setlength{\topsep}{0pt}
     \setlength{\partopsep}{0pt}
     \setlength{\leftmargin}{1.95em}
     \setlength{\labelwidth}{1.5em}
     \setlength{\labelsep}{0.5em} } }

\newcounter{Lcount}
\newcommand{\squishlist}{
    \begin{list}{\arabic{Lcount}. }
   { \usecounter{Lcount}
        \setlength{\itemsep}{0pt}
        \setlength{\parsep}{3pt}
        \setlength{\topsep}{0pt}
        \setlength{\partopsep}{0pt}
        \setlength{\leftmargin}{2em}
        \setlength{\labelwidth}{1.5em}
        \setlength{\labelsep}{0.5em} } }

\newcommand{\squishend}{\end{list}}

\newcommand{\bit}{\begin{compactitem}}
\newcommand{\eit}{\end{compactitem}}
\newcommand{\ben}{\begin{compactenum}}
\newcommand{\een}{\end{compactenum}}

\newcommand{\eg}{\textit{e}.\textit{g}.,\xspace}
\newcommand{\ie}{\textit{i}.\textit{e}.,\xspace}

\definecolor{dkgreen}{rgb}{0,0.6,0}

\def\Snospace~{\S{}}

\title{\paperTitle} 
\titlenote{The first two authors contributed equally to this paper}

\begin{document}

 \author{Renzhi Wu*$^1$, Pramod Chunduri*$^1$, Dristi J Shah$^1$, Ashmitha Julius Aravind$^1$, Ali Payani$^2$, Xu Chu$^1$, Joy Arulraj$^1$, Kexin Rong$^1$}
 \affiliation{%
   \institution{$^1$Georgia Institute of Technology, $^2$Cisco}
 }
 \email{{renzhiwu@, pramodc@,dshah371@,asjula@, xu.chu@cc., arulraj@, krong@}gatech.edu, apayani@cisco.com}








\newcommand{\kr}[1]{\textcolor{blue}{\textbf{KR: } #1}}

%
\begin{abstract}
In this paper, we will present \sys, a video database management system (VDBMS) for retrieving video moments with a sketch-based query interface. 
This novel interface allows users to specify object trajectory events with simple mouse drag-and-drop operations. 
Users can use trajectories of single objects as building blocks to compose complex events. 
Using a pre-trained model that encodes trajectory similarity, \sys achieves zero-shot video moments retrieval by performing similarity searches over the video to identify clips that are the most similar to the visual query. 
In this demonstration, we introduce the graphic user interface of \sys and detail its functionalities and interaction mechanisms. We also demonstrate the end-to-end usage of \sys from query composition to video moments retrieval using real-world scenarios. 

\end{abstract}

\maketitle

\pagestyle{\vldbpagestyle}
\begingroup\small\noindent\raggedright\textbf{PVLDB Reference Format:}\\
\vldbauthors. \vldbtitle. PVLDB, \vldbvolume(\vldbissue): \vldbpages, \vldbyear.\\
\href{https://doi.org/\vldbdoi}{doi:\vldbdoi}
\endgroup
\begingroup
\renewcommand\thefootnote{}\footnote{\noindent
This work is licensed under the Creative Commons BY-NC-ND 4.0 International License. Visit \url{https://creativecommons.org/licenses/by-nc-nd/4.0/} to view a copy of this license. For any use beyond those covered by this license, obtain permission by emailing \href{mailto:info@vldb.org}{info@vldb.org}. Copyright is held by the owner/author(s). Publication rights licensed to the VLDB Endowment. \\
\raggedright Proceedings of the VLDB Endowment, Vol. \vldbvolume, No. \vldbissue\ %
ISSN 2150-8097. \\
\href{https://doi.org/\vldbdoi}{doi:\vldbdoi} \\
}\addtocounter{footnote}{-1}\endgroup

\ifdefempty{\vldbavailabilityurl}{}{
\vspace{.3cm}
\begingroup\small\noindent\raggedright\textbf{PVLDB Artifact Availability:}\\
The source code, data, and/or other artifacts have been made available at \url{\vldbavailabilityurl}.
\endgroup
}

\section{Introduction}
\label{sec:introduction}
Video moment retrieval is an important task in video analytics whose goal is to search for target moments (sequences of frames) within a video. 
%
This task has numerous applications in traffic surveillance, sports analytics, and autonomous driving.

\begin{figure}[t]
  \centering  
   \subfloat[Nearby car, acute angle, turning top.]{\label{fig:left1}{\includegraphics[width=0.3\linewidth]{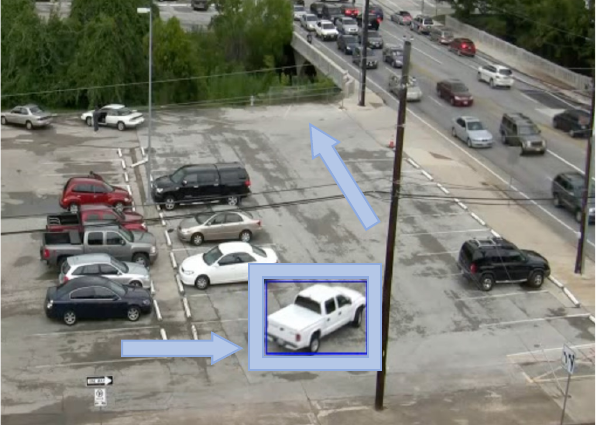}}} \hfill
       \subfloat[Distant car, acute angle, turning top.]{\label{fig:left2}{\includegraphics[width=0.3\linewidth]{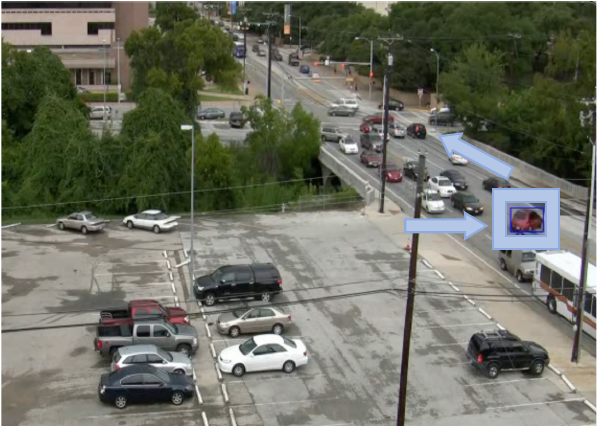}}} \hfill
    \subfloat[Distant car, obtuse angle, turning left.]{\label{fig:left3}{\includegraphics[width=0.3\linewidth]{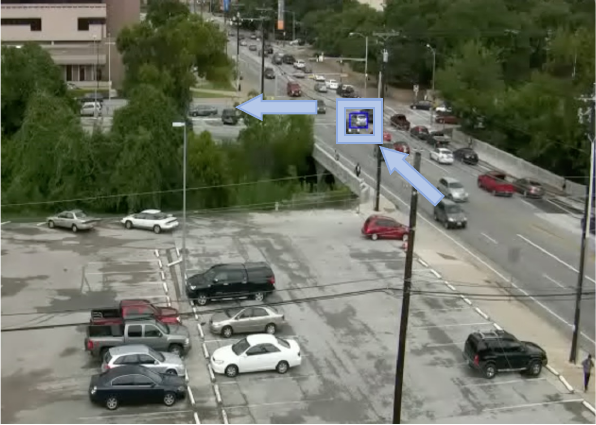}}} \hfill
    \vspace{-2mm}
    \caption{
      \textbf{
      Diverse left-turn behaviors in a real-world traffic surveillance video stream. 
      }
    }
  \label{fig:left-turn-example}
\end{figure}

For example, transport researchers are interested in retrieving different instances of left-turning vehicles from surveillance video streams to analyze driving behaviors and improve traffic safety~\cite{abdeljaber2020analysis}.
However, accurately detecting "left turn" motions in diverse, real-world videos can be quite challenging. 
Consider the illustrative video clips from a parking lot surveillance camera in \cref{fig:left-turn-example}. 
In \cref{fig:left1}, a car begins driving towards the right side of the screen and makes a left turn towards the top, while in \cref{fig:left3}, another car starts driving towards the top of the screen and makes a left turn towards the left side. 
Furthermore, due to the camera's varying position and angle relative to vehicles, the turning angles might appear different on camera: the car in \cref{fig:left2} has an acute turning angle, while the car in \cref{fig:left3} has an obtuse angle.
Ideally, the left turn query should capture all types of left-turn events, irrespective of the vehicle's initial direction or turning angle. 
Although the camera is stationary in this example, it could still subject to movements caused by environmental factors like wind or vibrations; other video streams, such as those in sports, often feature moving and panning cameras. These further complicates the identification of relevant events.

\PP{Limitations of Current Approaches} 
The two main types of query interfaces for video moment retrieval, natural language-based and SQL-based, suffer from limitations in generalizability or ease-of-use and cannot adequately address the above challenges.
\squishlist
\item Natural language-based interfaces retrieve target video clips based on user-specified text (\eg ``Car making a left turn") and are popular within the machine learning community~\cite{liu2023survey}.
%
These methods are easy to use for non-experts, and are typically implemented by training end-to-end deep learning models that map text to raw video frames~\cite{zhang2020learning, zhang2021multi}.
A key limitation of these interfaces is their reliance on large training data to achieve accurate retrieval~\cite{anne2017localizing},
which limits their application outside the original training contexts.
\item SQL-based query interfaces, predominantly developed within the data management community~\cite{kang2018blazeit, bastani2020miris, xu2022eva}, support rule-based selection of clips using SQL-like syntax. 
They are often built upon low-level primitives extracted by pre-trained models, such as pre-trained object detectors~\cite{kang2018blazeit,xu2022eva}, object tracking models~\cite{kang2018blazeit,xu2022eva}, or scene graph extraction models~\cite{chen2022spatial}.
The main advantage of SQL-based interfaces is their ability to \emph{generalize} across different datasets and video domains with few or no labeled examples, thanks to the robust performance of pre-trained models~\cite{shao2019objects365}.
However, SQL-base interfaces require considerable query specification time from users because translating a semantically meaningful event (\eg left turns) into SQL-like rules on top of low-level primitives (\eg location and angle of bounding boxes) can be challenging. 
\squishend

\PP{Video Moment Querying with a sketch-based interface}
To address the above limitations, in our recent work~\cite{sketchqlpaper}, we developed \sys, a video database management system (VDBMS) for offline, exploratory video moment retrieval that is both easy to use and generalizes well across multiple video moment datasets.
To improve ease-of-use, \sys features a \textit{visual query interface} that enables users to sketch complex visual queries through intuitive drag-and-drop actions.
By leveraging a human's inherent ability to capture complex events via sketches, \sys improves the usability and expressivity of query specifications for non-expert users. 
To improve generalizability, \sys operates on object-tracking primitives that are reliably extracted across various datasets using pre-trained models~\cite{shao2019objects365}.
The sketch-based query is executed by comparing the query trajectories provided by the user to object bounding box trajectories extracted using pre-trained object trackers.
We developed a transformer-based neural network model that learns a similarity measure between trajectories that is robust to differences in camera perspectives and movements. 
%
\sys trains the model on a diverse dataset generated with a novel simulator, that enhances its accuracy across a wide array of datasets and queries.

In this paper, we introduce the user interface of \sys. We demonstrate the main functionalities of \sys (composing visual queries and retrieving similar clips) with end2end scenarios. 

\section{System Overview}
\label{sec:overview}

\begin{figure*}[htb!]
	\centering
	\includegraphics[width=0.6\linewidth]{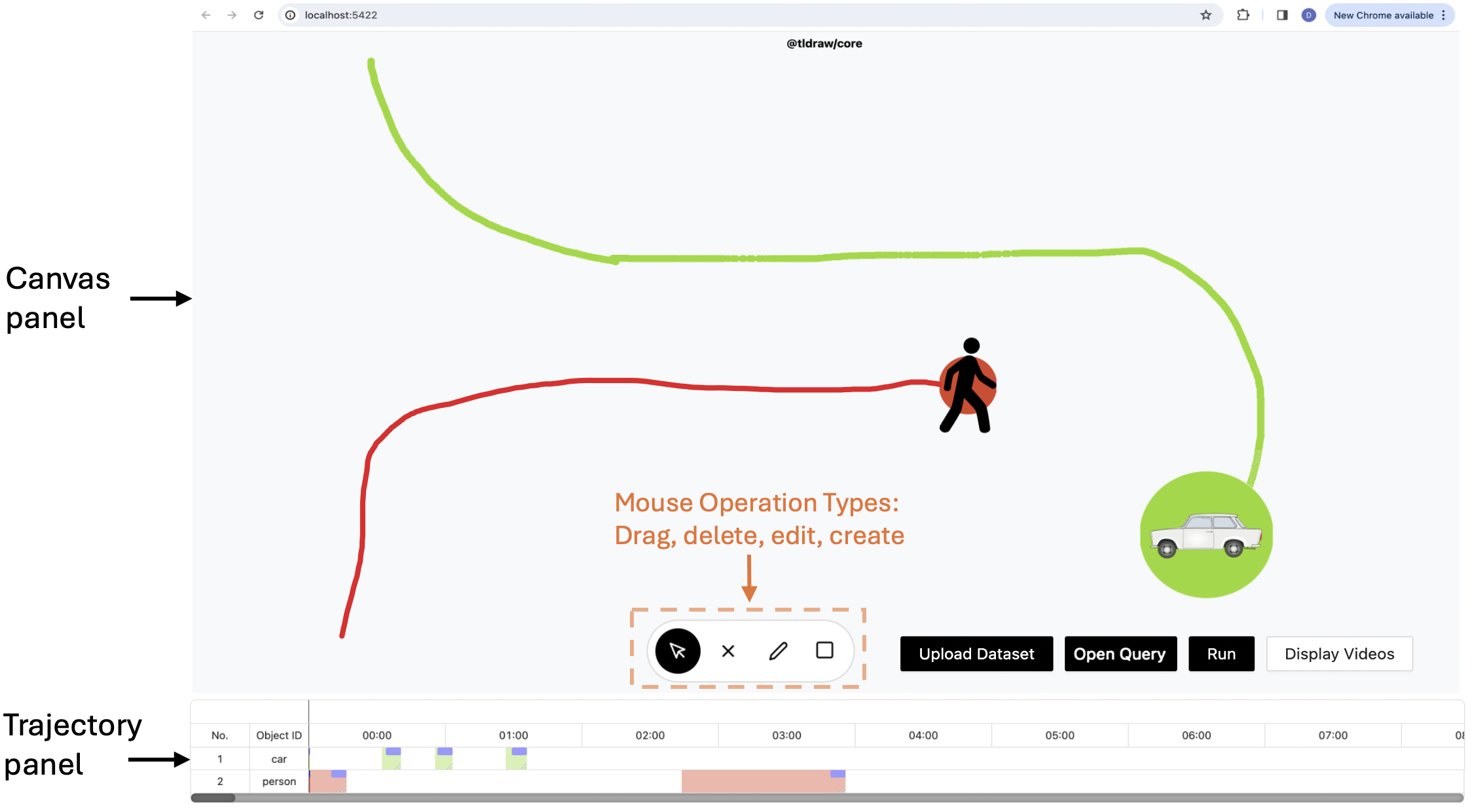}
	\caption{ User interface of \sys.}
	\label{fig:system-interface} 
\end{figure*}

\sys consists of three key components: (1) \sketcher features a \textit{visual query interface} that enables users to sketch complex queries through simple drag-and-drop actions, (2) \matcher compares the query trajectories provided by the user to object bounding box trajectories extracted using pre-trained object trackers and (3) \tuner is an optional component that incorporates explicit user feedback when provided to improve the retrieval quality.

\sys is designed to operate on top of per-frame object bounding boxes rather than raw pixels, similar to \miris~\cite{bastani2020miris} and STAR Retrieval~\cite{chen2022spatial}.
Bounding box sequences for objects across frames are obtained in a preprocessing step using pre-trained object trackers~\cite{zhang2022bytetrack} without dataset-specific retraining.

%

\subsection{\sketcher: Composing visual queries}
\label{section:interface}
The \sketcher is the query interface of \sys, and it has two major components:
(a) \textit{The Canvas}. This is a whiteboard where users can place and drag objects to compose clips (\cref{fig:system-interface}, top).
(b) \textit{The Trajectory Panel}. This is a panel where users can adjust multiple trajectories of the same object and align the trajectories of different objects (\cref{fig:system-interface}, bottom). 
The \sketcher is developed on top of a popular canvas library \texttt{tldraw}\footnote{\url{https://github.com/tldraw/tldraw}}. 
We detail the specific functionalities of \sketcher in the following. 
%

When using the canvas panel, the user can select one of four types of mouse operations including: drag, delete, edit, and create by clicking the corresponding buttons (shown in the pink dashed box in~\cref{fig:system-interface}). 
The four operations enable the following functionalities:

\noindent(1) \textbf{Object Creation}. This action allows users to select an object type and place the object on the canvas.
When the user clicks on the "square" icon to select the mouse operation type as "create", 
an input box pops up for the user to specify the object type.
Currently, about eighty common object types (\eg car, person) are supported. The user can also set a generic type \emph{Any} that represents any types of objects. After setting the object type, the user can place one or more objects on the canvas. 

\noindent(2) \textbf{Object Deletion and Editing.} The user is allowed to delete or edit an object on the canvas. When the user clicks on the "cross" icon, the mouse enters deletion mode and the user can click on an object to delete it. Similarly, when the user clicks on the "pencil" icon, the user can click on an object to change its object type. 

\noindent(3) \textbf{Trajectory Creation with Drag and Drop}. 
The user can create trajectories when the mouse is in "Drag" mode which activates when the user clicks on the "cursor" icon. 
In this mode, the user can move an object on the Canvas via mouse drag-and-drop operations, and
all the coordinates of the movements are automatically recorded. 
Each drag-and-drop operation is represented as a box in the trajectory panel. 
For example, the car movement in~\cref{fig:system-interface} is created by three drag-and-drop operations: left turn, going straight, and right turn, so there are three boxes shown in the trajectory panel for the car object. This abstraction allows users to use each drag-and-drop operation as a building block to compose complex motions for an object, even enabling complex events involving multiple objects. 

%

The Trajectory Panel is similar to the soundtrack panel in existing audio editing tools. 
The trajectory panel allows for the following functionalities:

\noindent(1) \textbf{Trajectory Editing}. The user can delete the boxes in the trajectory panel for each object to remove unwanted trajectories. The user can resize the box to make it shorter or longer to speed up or slow down the corresponding trajectory. The user can also rearrange the boxes to change the order of the trajectories. For example, in Figure~\ref{fig:system-interface}, the car's initial trajectory is a left turn, followed by straight motion, and then a right turn. By reordering the three boxes, users can edit the event to depict a different sequence, such as a left turn, a right turn, and then straight motion.

\noindent(2) \textbf{Trajectory Coordination Among Multiple Objects.} When creating events with multiple objects, we need to coordinate the timing of the trajectories. For example, in the event depicted in~\cref{fig:system-interface}, the motion that the car goes right happens first, and after a long time the person goes right. To create a synchronized event in which the person and the car go right at the same time, we can move the second box of the person object in the trajectory panel to align with the second box of the car object. 

%
%

Apart from the canvas panel and the trajectory panel, the interface also provides a few buttons in the bottom right corner as shown in~\cref{fig:system-interface}. The buttons provide the following functionalities: (1) \textit{Dataset Uploading.} When the user clicks on the button "Upload Dataset", a window pops out allowing the user to select a video file to upload. After a dataset is uploaded, future queries will be executed using this dataset. (2) \textit{Query replay.} When the user clicks on the "Open Query" button, a window pops out displaying an video that animates the defined event, so the user can check the query holistically, and optionally go back to editing the query. 
(3) \textit{Query execution.} When the user clicks on the button "Run", the visual query is sent to backend to be executed, where \matcher is invoked to find similar video clips. (4) \textit{Display query results.} When the user clicks on the button "Display Videos", a window pops out listing the found similar video clips. 

\subsection{\matcher: Identifying Similar Clips}
\label{sec:simsearch}
The \matcher is the backend of \sys. We discuss it briefly here, and more details can be found in our research paper~\cite{sketchqlpaper}. 

The \matcher identifies video clips $C_V$ that are most similar to a given visual query $C_Q$ through sliding-window similarity search.
The core challenge is defining the similarity function $sim(C_Q, C_V)$ that is robust to camera angles and noises.
We developed a pre-trained model that encodes the bounding box trajectories in $C_Q$ or $C_V$ as embeddings, and then similarity is measured as the cosine similarity between the embeddings. 
The model architecture is a transformer that encodes the the sequences of bounding box trajectories of multiple objects into one embedding vector. 

To train the embedding model, we propose synthesizing labeled training data using a custom trajectory simulator, inspired by the wide adoption of simulators in generating training data for autonomous driving models.
The simulator operates on top of the bounding box abstraction, enabling it to synthesize trajectories across video domains. 
The high-level idea of our simulator is to generate motions in a 3D space and create 2D video clips by recording the event from virtual cameras placed at random locations in the 3D space. Intuitively, 2D video clips from the different cameras of the same 3D clip are positive (similar) examples, and 2D video clips from different 3D clips are negative (dissimilar) examples.

\sys also has an optional \tuner component that can adapt the learned similarity measure according to user feedback.
Since \sys provides decent results without needing user feedback, in this paper we focus on demonstrating the zero-shot retrieval capability of \sys and omit the discussion on \tuner. 



\section{Demonstration}
\label{sec:demonstration}

\begin{figure*}[ht!]
  \centering  
   \subfloat[Upload Dataset.]{\label{fig:scenario-upload}{\includegraphics[width=0.4\linewidth]{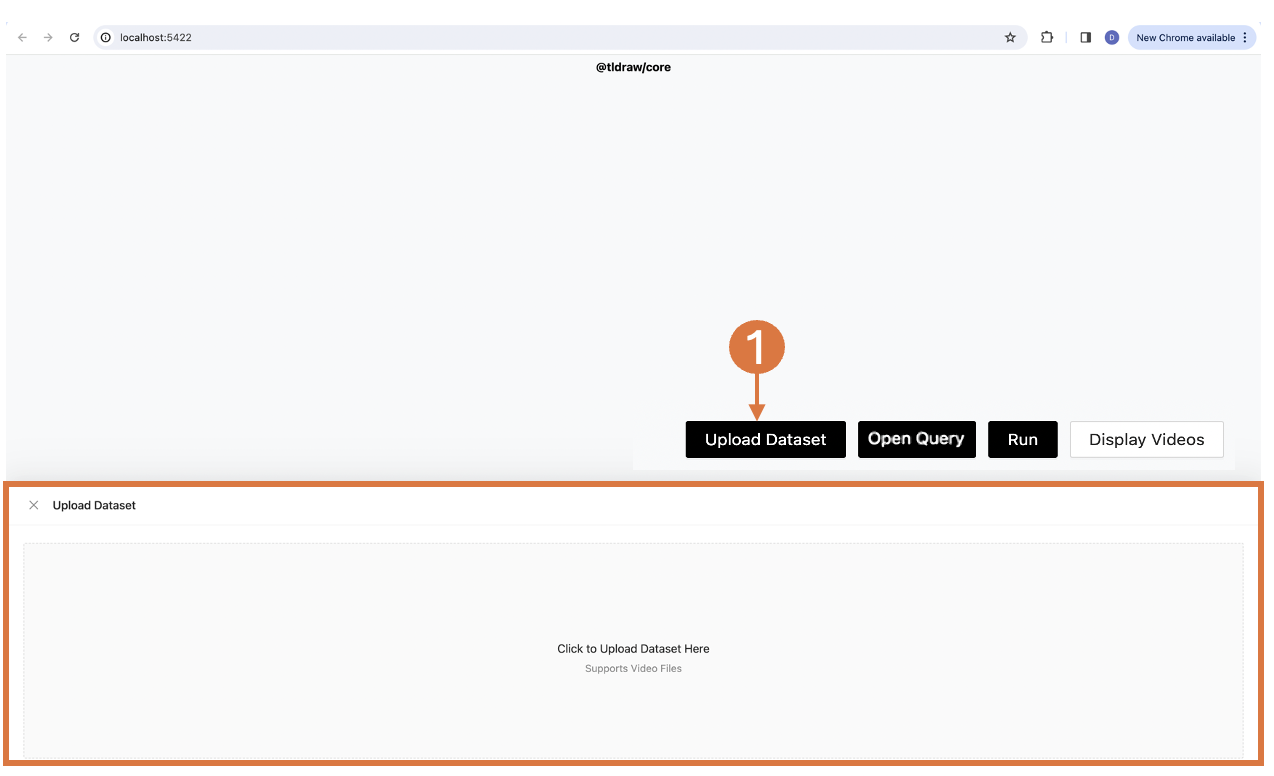}}} \hspace{2em}
    \subfloat[Create object.]{\label{fig:scenario-create}{\includegraphics[width=0.4\linewidth]{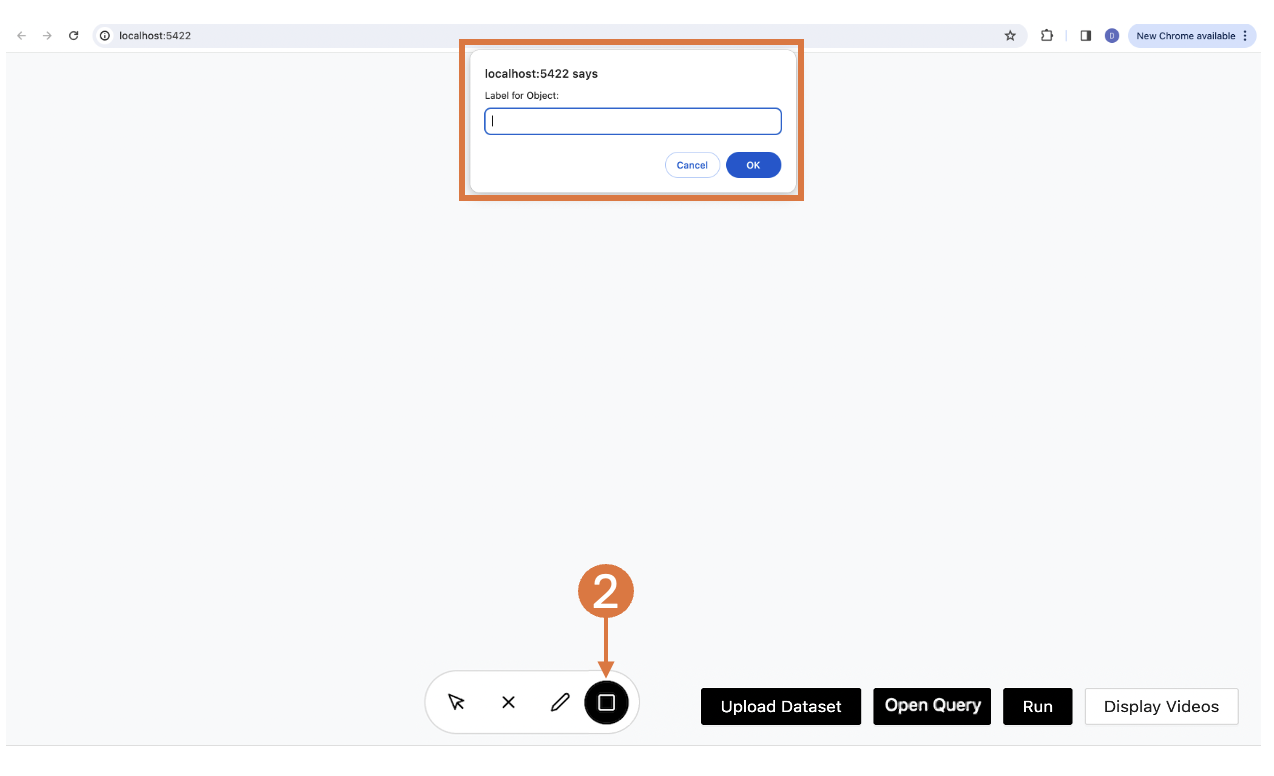}}}\\
    \subfloat[Create and Edit Trajectory.]{\label{fig:scenario-trajectory}{\includegraphics[width=0.4\linewidth]{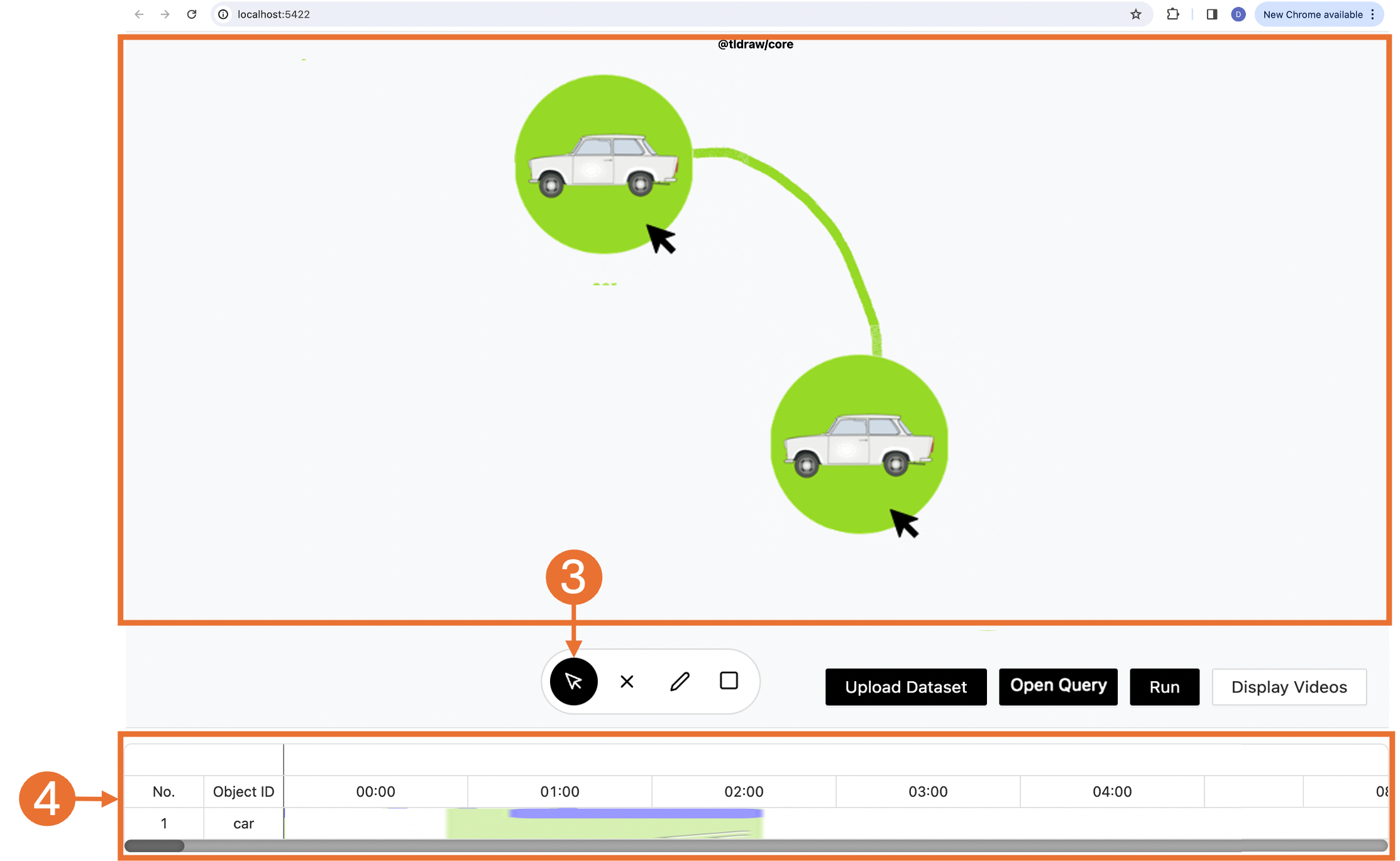}}} 
    \hspace{2em}
    \subfloat[Run query and display results.]{\label{fig:scenario-run}{\includegraphics[width=0.4\linewidth]{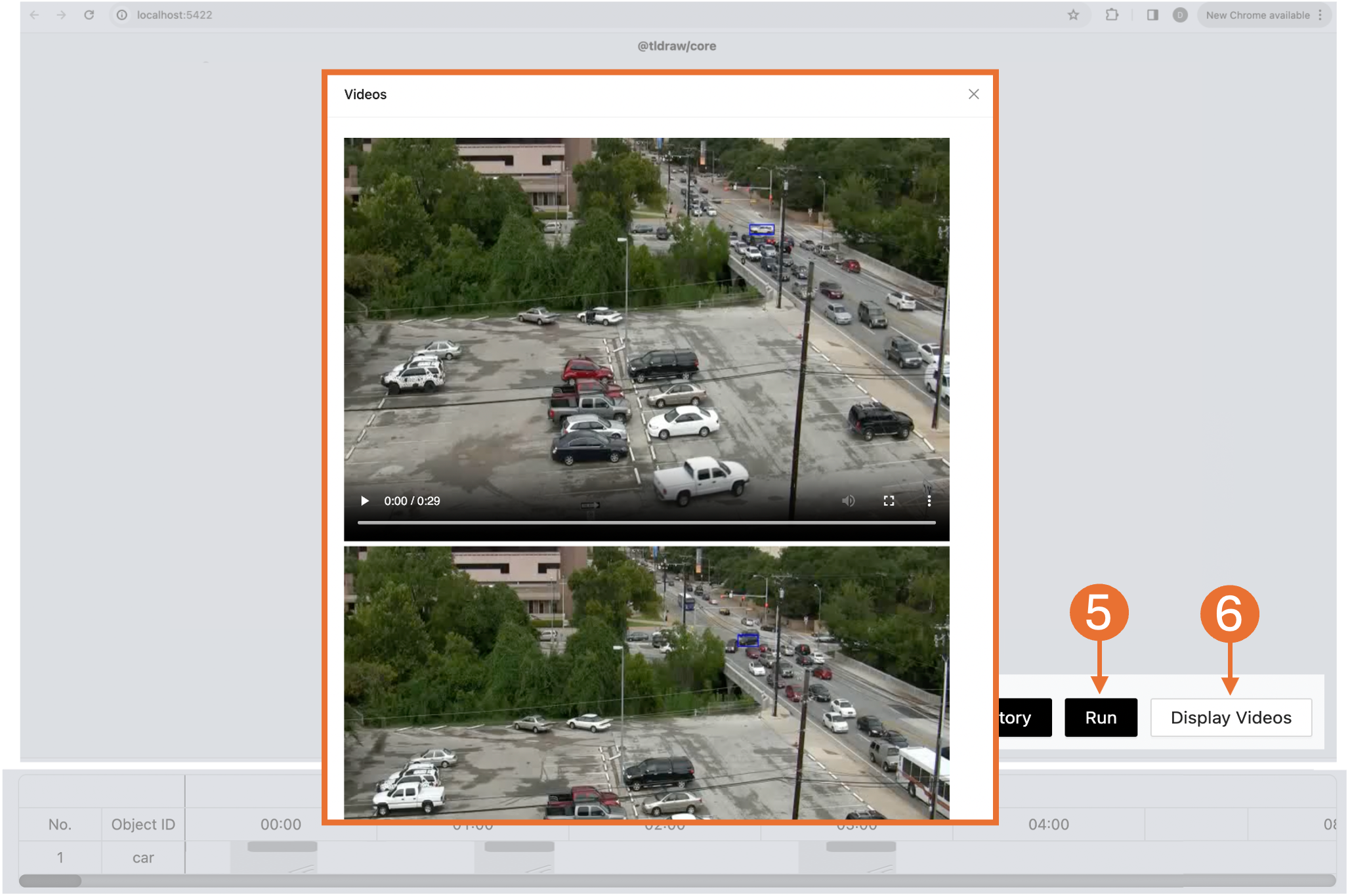}}} 
    \caption{
      \textbf{\sys end-to-end usage demonstration}
    }
  \label{fig:usage-scenarios}
  \vspace{-3mm}
\end{figure*}

We demonstrate the usage of \sys with an end-to-end scenario. We use a real-world traffic surveillance dataset~\cite{oh2011large} that is used extensively in the computer vision community. We consider two common queries: (Q1) a car making a left turn and (Q2) car and person moving perpendicularly to each other. 
Q1 only has one object, while Q2 involves two objects. The overall pipeline of Q2 is similar to that of Q1; it only differs in how to coordinate the two objects. 
Therefore, we first illustrate the overall pipeline with Q1, and then we discuss how to coordinate multiple objects with Q2. 

\subsection{End-to-end Demo with Q1}
\label{ssec:q1-demo}
The end-to-end usage of \sys typically involves six steps and is illustrated in Figure~\ref{fig:usage-scenarios}. 

\noindent\textbf{Step 1: Uploading dataset and initialization.} The user uploads a video by clicking the "Upload Dataset" button and selecting a video file in the pop-up window. After this, \sys initializes by extracting the object tracking primitives for the dataset. 

\noindent\textbf{Step 2: Object Creation.}
The user clicks on the creation icon, \ie the "square" icon, and sets the object type to "Car" in the pop-up input window. After that, the user clicks on the canvas to place the "Car" object. 

\noindent\textbf{Step 3: Trajectory Creation.}
The user clicks the "cursor" icon to enable the mouse's "Drag" mode. Then, the user drags the "Car" object on the canvas to make a left turn. The user drops the "Car" object when the left turn finishes. 

\noindent\textbf{Step 4: Trajectory Editing.}
The user clicks on the "Open Query" button to replay the event just created. The user can edit the trajectory if he/she is not satisfied. For example, 
the user may delete the trajectory from the Trajectory Panel and then drag the object to create a new trajectory. If the user is satisfied with the shape of the trajectory but hopes to let the car make a left turn faster, the user can stretch the corresponding box on the Trajectory Panel to shorten the box. 

\noindent\textbf{Step 5: Query Execution.}
The user clicks on the "Run" button. The visual query is then sent to the backend to be executed. 

\noindent\textbf{Step 6: Checking the Found Video Clips.}
After query execution, the user can click "Display Videos" to see the list of similar video clips found by the \matcher sorted by their query similarity scores. 

\subsection{Multi-object Event Query Demo with Q2}
We demonstrate how to create queries with multi-object events with a simple query, car, and person moving perpendicularly to each other. Step 1, Step 5, and Step 6 are the same as in Section~\ref{ssec:q1-demo}, and we explain Step 2, Step 3, and Step 4. 

\noindent\textbf{Step 2 (Multi-object): Object Creation.} The user follows Step 2 in Section~\ref{ssec:q1-demo} to create one "Car" object and one "Person" object. 

\noindent\textbf{Step 3 (Multi-object): Trajectory Creation.} The user clicks on the "cursor" icon to enable the mouse's "Drag" mode. The user drags the "Person" object to move horizontally. After that, the user drags the "Car" object to move vertically. With this creation, the "Person" object moves first, and the "Car" object moves after. To make them move simultaneously, next we coordinate their movements. 

\begin{figure}[H]
\vspace{-3mm}
  \centering \includegraphics[width=0.95\columnwidth]{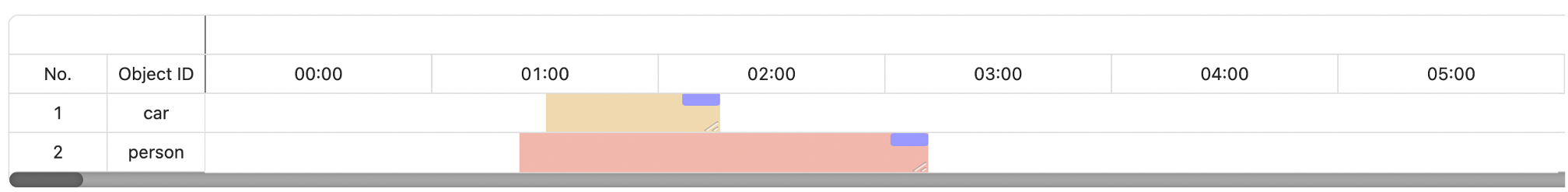}
    \caption{
      \textbf{
      Trajectory panel for multi-object query Q2 after Step 4.
      }
    }
  \label{fig:multi-object-trajectory}
  \vspace{-3mm}
\end{figure}

\noindent\textbf{Step 4 (Multi-object): Trajectory Editing.} 
In the Trajectory Panel, the user drags the box representing the trajectory of the "Car" object to the left to ensure it synchronizes with the box representing the trajectory of the "Person" object. Figure~\ref{fig:multi-object-trajectory} shows the Trajectory Panel after synchronization (originally the box for the "Car" object is on the right side of the box for the "Person" object). 

\section{Conclusion}
In this paper, we demonstrated \sys, a video database management system (VDBMS) for retrieving video moments with a visual query interface.  
\sys's interface allows query specification with simple mouse drag-and-drop operations. 
The interface also supports creating query with multiple objects.  

\noindent\textbf{Acknowledgements.} This work is supported by NSF under grant IIS-2335881 and IIS-2238431.

\bibliographystyle{ACM-Reference-Format}
\bibliography{main}


\begin{thebibliography}{13}


\ifx \showCODEN    \undefined \def \showCODEN     #1{\unskip}     \fi
\ifx \showDOI      \undefined \def \showDOI       #1{#1}\fi
\ifx \showISBNx    \undefined \def \showISBNx     #1{\unskip}     \fi
\ifx \showISBNxiii \undefined \def \showISBNxiii  #1{\unskip}     \fi
\ifx \showISSN     \undefined \def \showISSN      #1{\unskip}     \fi
\ifx \showLCCN     \undefined \def \showLCCN      #1{\unskip}     \fi
\ifx \shownote     \undefined \def \shownote      #1{#1}          \fi
\ifx \showarticletitle \undefined \def \showarticletitle #1{#1}   \fi
\ifx \showURL      \undefined \def \showURL       {\relax}        \fi
\providecommand\bibfield[2]{#2}
\providecommand\bibinfo[2]{#2}
\providecommand\natexlab[1]{#1}
\providecommand\showeprint[2][]{arXiv:#2}

\bibitem[\protect\citeauthoryear{Abdeljaber, Younis, and Alhajyaseen}{Abdeljaber et~al\mbox{.}}{2020}]%
        {abdeljaber2020analysis}
\bibfield{author}{\bibinfo{person}{Osama Abdeljaber}, \bibinfo{person}{Adel Younis}, {and} \bibinfo{person}{Wael Alhajyaseen}.} \bibinfo{year}{2020}\natexlab{}.
\newblock \showarticletitle{Analysis of the trajectories of left-turning vehicles at signalized intersections}.
\newblock \bibinfo{journal}{\emph{Transportation research procedia}}  \bibinfo{volume}{48} (\bibinfo{year}{2020}), \bibinfo{pages}{1288--1295}.
\newblock


\bibitem[\protect\citeauthoryear{Anne~Hendricks, Wang, Shechtman, Sivic, Darrell, and Russell}{Anne~Hendricks et~al\mbox{.}}{2017}]%
        {anne2017localizing}
\bibfield{author}{\bibinfo{person}{Lisa Anne~Hendricks}, \bibinfo{person}{Oliver Wang}, \bibinfo{person}{Eli Shechtman}, \bibinfo{person}{Josef Sivic}, \bibinfo{person}{Trevor Darrell}, {and} \bibinfo{person}{Bryan Russell}.} \bibinfo{year}{2017}\natexlab{}.
\newblock \showarticletitle{Localizing moments in video with natural language}. In \bibinfo{booktitle}{\emph{Proceedings of the IEEE international conference on computer vision}}. \bibinfo{pages}{5803--5812}.
\newblock


\bibitem[\protect\citeauthoryear{Bastani, He, Balasingam, Gopalakrishnan, Alizadeh, Balakrishnan, Cafarella, Kraska, and Madden}{Bastani et~al\mbox{.}}{2020}]%
        {bastani2020miris}
\bibfield{author}{\bibinfo{person}{Favyen Bastani}, \bibinfo{person}{Songtao He}, \bibinfo{person}{Arjun Balasingam}, \bibinfo{person}{Karthik Gopalakrishnan}, \bibinfo{person}{Mohammad Alizadeh}, \bibinfo{person}{Hari Balakrishnan}, \bibinfo{person}{Michael Cafarella}, \bibinfo{person}{Tim Kraska}, {and} \bibinfo{person}{Sam Madden}.} \bibinfo{year}{2020}\natexlab{}.
\newblock \showarticletitle{Miris: Fast object track queries in video}. In \bibinfo{booktitle}{\emph{Proceedings of the 2020 ACM SIGMOD International Conference on Management of Data}}. \bibinfo{pages}{1907--1921}.
\newblock


\bibitem[\protect\citeauthoryear{Chen, Koudas, Yu, and Yu}{Chen et~al\mbox{.}}{2022}]%
        {chen2022spatial}
\bibfield{author}{\bibinfo{person}{Yueting Chen}, \bibinfo{person}{Nick Koudas}, \bibinfo{person}{Xiaohui Yu}, {and} \bibinfo{person}{Ziqiang Yu}.} \bibinfo{year}{2022}\natexlab{}.
\newblock \showarticletitle{Spatial and temporal constrained ranked retrieval over videos}.
\newblock \bibinfo{journal}{\emph{Proceedings of the VLDB Endowment}} \bibinfo{volume}{15}, \bibinfo{number}{11} (\bibinfo{year}{2022}), \bibinfo{pages}{3226--3239}.
\newblock


\bibitem[\protect\citeauthoryear{Kang, Bailis, and Zaharia}{Kang et~al\mbox{.}}{2018}]%
        {kang2018blazeit}
\bibfield{author}{\bibinfo{person}{Daniel Kang}, \bibinfo{person}{Peter Bailis}, {and} \bibinfo{person}{Matei Zaharia}.} \bibinfo{year}{2018}\natexlab{}.
\newblock \showarticletitle{BlazeIt: optimizing declarative aggregation and limit queries for neural network-based video analytics}.
\newblock \bibinfo{journal}{\emph{arXiv preprint arXiv:1805.01046}} (\bibinfo{year}{2018}).
\newblock


\bibitem[\protect\citeauthoryear{Liu, Nie, Wang, Wang, and Rui}{Liu et~al\mbox{.}}{2023}]%
        {liu2023survey}
\bibfield{author}{\bibinfo{person}{Meng Liu}, \bibinfo{person}{Liqiang Nie}, \bibinfo{person}{Yunxiao Wang}, \bibinfo{person}{Meng Wang}, {and} \bibinfo{person}{Yong Rui}.} \bibinfo{year}{2023}\natexlab{}.
\newblock \showarticletitle{A survey on video moment localization}.
\newblock \bibinfo{journal}{\emph{Comput. Surveys}} \bibinfo{volume}{55}, \bibinfo{number}{9} (\bibinfo{year}{2023}), \bibinfo{pages}{1--37}.
\newblock


\bibitem[\protect\citeauthoryear{Oh, Hoogs, Perera, Cuntoor, Chen, Lee, Mukherjee, Aggarwal, Lee, Davis, et~al\mbox{.}}{Oh et~al\mbox{.}}{2011}]%
        {oh2011large}
\bibfield{author}{\bibinfo{person}{Sangmin Oh}, \bibinfo{person}{Anthony Hoogs}, \bibinfo{person}{Amitha Perera}, \bibinfo{person}{Naresh Cuntoor}, \bibinfo{person}{Chia-Chih Chen}, \bibinfo{person}{Jong~Taek Lee}, \bibinfo{person}{Saurajit Mukherjee}, \bibinfo{person}{JK Aggarwal}, \bibinfo{person}{Hyungtae Lee}, \bibinfo{person}{Larry Davis}, {et~al\mbox{.}}} \bibinfo{year}{2011}\natexlab{}.
\newblock \showarticletitle{A large-scale benchmark dataset for event recognition in surveillance video}. In \bibinfo{booktitle}{\emph{CVPR 2011}}. IEEE, \bibinfo{pages}{3153--3160}.
\newblock


\bibitem[\protect\citeauthoryear{Shao, Li, Zhang, Peng, Yu, Zhang, Li, and Sun}{Shao et~al\mbox{.}}{2019}]%
        {shao2019objects365}
\bibfield{author}{\bibinfo{person}{Shuai Shao}, \bibinfo{person}{Zeming Li}, \bibinfo{person}{Tianyuan Zhang}, \bibinfo{person}{Chao Peng}, \bibinfo{person}{Gang Yu}, \bibinfo{person}{Xiangyu Zhang}, \bibinfo{person}{Jing Li}, {and} \bibinfo{person}{Jian Sun}.} \bibinfo{year}{2019}\natexlab{}.
\newblock \showarticletitle{Objects365: A large-scale, high-quality dataset for object detection}. In \bibinfo{booktitle}{\emph{Proceedings of the IEEE/CVF international conference on computer vision}}. \bibinfo{pages}{8430--8439}.
\newblock


\bibitem[\protect\citeauthoryear{Wu, Chunduri, Payani, Chu, Arulraj, and Rong}{Wu et~al\mbox{.}}{2025}]%
        {sketchqlpaper}
\bibfield{author}{\bibinfo{person}{Renzhi Wu}, \bibinfo{person}{Pramod Chunduri}, \bibinfo{person}{Ali Payani}, \bibinfo{person}{Xu Chu}, \bibinfo{person}{Joy Arulraj}, {and} \bibinfo{person}{Kexin Rong}.} \bibinfo{year}{2025}\natexlab{}.
\newblock \showarticletitle{SketchQL: Video Moment Querying with a Visual Query Interface}. In \bibinfo{booktitle}{\emph{Proceedings of the 2025 ACM SIGMOD International Conference on Management of Data}}.
\newblock


\bibitem[\protect\citeauthoryear{Xu, Kakkar, Arulraj, and Ramachandran}{Xu et~al\mbox{.}}{2022}]%
        {xu2022eva}
\bibfield{author}{\bibinfo{person}{Zhuangdi Xu}, \bibinfo{person}{Gaurav~Tarlok Kakkar}, \bibinfo{person}{Joy Arulraj}, {and} \bibinfo{person}{Umakishore Ramachandran}.} \bibinfo{year}{2022}\natexlab{}.
\newblock \showarticletitle{EVA: A symbolic approach to accelerating exploratory video analytics with materialized views}. In \bibinfo{booktitle}{\emph{Proceedings of the 2022 International Conference on Management of Data}}. \bibinfo{pages}{602--616}.
\newblock


\bibitem[\protect\citeauthoryear{Zhang, Peng, Fu, Lu, and Luo}{Zhang et~al\mbox{.}}{2021}]%
        {zhang2021multi}
\bibfield{author}{\bibinfo{person}{Songyang Zhang}, \bibinfo{person}{Houwen Peng}, \bibinfo{person}{Jianlong Fu}, \bibinfo{person}{Yijuan Lu}, {and} \bibinfo{person}{Jiebo Luo}.} \bibinfo{year}{2021}\natexlab{}.
\newblock \showarticletitle{Multi-scale 2d temporal adjacency networks for moment localization with natural language}.
\newblock \bibinfo{journal}{\emph{IEEE Transactions on Pattern Analysis and Machine Intelligence}} \bibinfo{volume}{44}, \bibinfo{number}{12} (\bibinfo{year}{2021}), \bibinfo{pages}{9073--9087}.
\newblock


\bibitem[\protect\citeauthoryear{Zhang, Peng, Fu, and Luo}{Zhang et~al\mbox{.}}{2020}]%
        {zhang2020learning}
\bibfield{author}{\bibinfo{person}{Songyang Zhang}, \bibinfo{person}{Houwen Peng}, \bibinfo{person}{Jianlong Fu}, {and} \bibinfo{person}{Jiebo Luo}.} \bibinfo{year}{2020}\natexlab{}.
\newblock \showarticletitle{Learning 2d temporal adjacent networks for moment localization with natural language}. In \bibinfo{booktitle}{\emph{Proceedings of the AAAI Conference on Artificial Intelligence}}, Vol.~\bibinfo{volume}{34}. \bibinfo{pages}{12870--12877}.
\newblock


\bibitem[\protect\citeauthoryear{Zhang, Sun, Jiang, Yu, Weng, Yuan, Luo, Liu, and Wang}{Zhang et~al\mbox{.}}{2022}]%
        {zhang2022bytetrack}
\bibfield{author}{\bibinfo{person}{Yifu Zhang}, \bibinfo{person}{Peize Sun}, \bibinfo{person}{Yi Jiang}, \bibinfo{person}{Dongdong Yu}, \bibinfo{person}{Fucheng Weng}, \bibinfo{person}{Zehuan Yuan}, \bibinfo{person}{Ping Luo}, \bibinfo{person}{Wenyu Liu}, {and} \bibinfo{person}{Xinggang Wang}.} \bibinfo{year}{2022}\natexlab{}.
\newblock \showarticletitle{Bytetrack: Multi-object tracking by associating every detection box}. In \bibinfo{booktitle}{\emph{ECCV}}. Springer, \bibinfo{pages}{1--21}.
\newblock


\end{thebibliography}

\end{document}